# Versatile and widely tunable mid-infrared erbium doped ZBLAN fiber laser

ORI HENDERSON-SAPIR,[1*] STUART D. JACKSON,[2] DAVID J. OTTAWAY[1]

[1]*Department of Physics and Institute for Photonics and Advanced Sensing (IPAS), The University of Adelaide, SA 5005, Australia*
[2]*MQ Photonics, Department of Engineering, Macquarie University, North Ryde, N.S.W. 2109, Australia*
*Corresponding author: ori.henderson-sapir@adelaide.edu.au*

**We report on a long wavelength emitting rare earth doped fiber laser with emission centered at 3.5 µm and tunable across 450 nm. The longest wavelength emission was 3.78 µm, which is the longest emission from a fiber laser operating at room temperature. In a simple optical arrangement employing dielectric mirrors for feedback, the laser was capable of emitting 1.45 W of near diffraction limited output power at 3.47 µm. These emission characteristics compliment the emission from quantum cascade lasers and demonstrate how all infrared dual wavelength pumping can be used to access high lying rare earth ion transitions that have previously relied on visible wavelength pumping.**

Advances in mid-infrared sensing have been predominantly associated with the development of narrow linewidth quantum cascade lasers [1-2], comb generators [3] and optical parametric oscillators [4]. At the shorter wavelength end between 2.5 µm and 6 µm, many compounds exhibit strong absorption features that can be used to identify species. Of particular interest is the functional group absorption region between 2.5 µm and 4 µm which offers opportunities for sensing and the precise and highly reproducible modification of industrial and biomedical materials [5, 6]. This region can also provide absorption features relevant to combination and overtone absorptions from the fingerprint region and benefits from the wide availability of fast and more sensitive photodiode detection.

Fiber lasers can generate very high powers in the near-infrared and their output has recently been pushed towards the mid-infrared [7]. After this review was written a Raman fiber laser produced 50mW at 3.34 µm [8]. Recently, we introduced an optical pumping method that allows the use of mature, well developed near infrared sources to excite high lying energy levels of rare earth ions [9]. In this work, we demonstrated that dual wavelength pumping (DWP) could significantly increase the efficiency of erbium doped ZBLAN glass fiber lasers that operate on the $^4F_{9/2} \rightarrow {^4I_{9/2}}$ transition at 3.5 µm [9]. Unlike previous reports of upconverted emission from high lying states that also employed excited state absorption for excitation of energetic states, our method allowed effective access to mid-infrared emitting transitions that previously have only been accessible using inconvenient visible wavelength pump sources. The concept is easily adaptable to other rare earth ions provided that *i)* relevant ground and excited state absorption features can be found that are commensurate with the emission from high power diode and fiber lasers and *ii)* that the intermediate energy level be sufficiently long lived to provide the "virtual ground state". The functional group region of the mid-infrared can now be accessed using moderate power and potentially high power fiber laser sources opening up the opportunity for the development of high brightness pump sources for applications in mid-infrared nonlinear optics and high peak power mid-infrared lasers. The recent report [10] of 1.5 W at a wavelength of 3.44 µm demonstrates the power scaling capability of the concept.

We now report a dual-wavelength-pumped grating-tuned $Er^{3+}$-doped ZBLAN laser that demonstrates the widest tuning range (450nm) of any rare earth doped laser. This laser can operate at 3780 nm which is the longest wavelength demonstrated by a fiber laser at room temperature. This was achieved using simple coated mirror optics and the implementation of greater pump power levels. We show a versatile implementation of the laser by converting the output to the $^4I_{11/2} \rightarrow {^4I_{13/2}}$ transition at a wavelength of 2.8 µm by simply turning the grating and switching off the second pump.

Figure 1 shows the fluorescence spectrum of the $^4F_{9/2} \rightarrow {^4I_{9/2}}$ transition in relation to the other primary mid-infrared emitters including the erbium ($^4I_{11/2} \rightarrow {^4I_{13/2}}$) that employs fluoride glass for the host. The full bandwidth of this transition is the broadest of any of the rare earth ion emitting from an optical fiber in the infrared and suggests opportunities for both broad tuning and ultrashort pulse emission. In Figure 1 the $^4F_{9/2} \rightarrow {^4I_{9/2}}$ transition is the only transition in a singly doped system that approaches truly four-level-laser-like behavior because its lower lasing transition is short lived (~5µs); the other transitions require co-doping with de-sensitizer ions to quench the lifetime of the lower laser level. The energy level diagram of the erbium ion is complicated because there are many energy-transfer processes due to the near degeneracies of the energy levels; the number grows for laser transitions that are located high above the ground state.

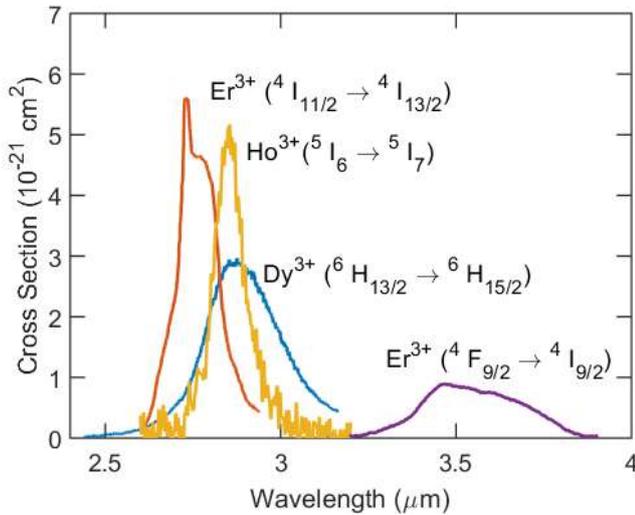

**Figure 1** – Measured emission cross sections of the rare-earth ions when doped into ZBLAN glass that have emission wavelengths longer than 2.5 μm.

Figure 2 shows schematically the energy level diagram of the lowest five levels of erbium ion, the two lasing transitions relevant to this work and the two pump wavelengths. In an earlier work [9] we showed that DWP is an efficient mean of generating emission from the 3.5 μm band transition of erbium. A population of ions is established in the metastable $^4I_{11/2}$ level by absorption of $P_1$ photons on the 974 nm to 985 nm ground state absorption band. The lasing cycle then involves excited ions absorbing $P_2$ pump photons at a wavelength of 1973 nm thus promoting them directly to the upper lasing level ($^4F_{9/2}$) which then emits mid-infrared light at 3.5 μm. The lower laser level ($^4I_{9/2}$) of the transition finally relaxes via rapid multi-phonon decay (MP) to the $^4I_{11/2}$ level. The lasing process relies on a ``virtual ground state'' that is created in the $^4I_{11/2}$ level because it has a lifetime of approximately 6 ms. The role of $P_1$ after the initial creation of population at the $^4I_{11/2}$ level is simply to replenish the ions that exit this cycle and return to the ground state. Of course lasing in the 2.8 μm band is simply achieved by pumping with only $P_1$.

A schematic of the experimental set-up is shown in Figure 3. The two pump beams are combined on a dichroic mirror before being injected by an aspheric lens through another dichroic mirror into the laser resonator. This dichroic mirror is highly reflective (HR) at 3.5 μm and transmits between 974 to 985 nm and 1973 nm. The mode-matching of each pump is tailored individually. The first pump is a commercial 30 W fiber-coupled laser diode (LIMO HLU30F200-980). The second pump laser is an in-house built 1973, $Tm^{3+}$-doped silicate fiber laser (TFL).

The $Er^{3+}$-doped ZBLAN fiber gain medium is butted against the HR surface of the dichroic mirror. The fiber is 2.8 m long and was manufactured by Le Verre Fluoré (France). It was double-clad fiber with a 16 μm diameter core, and 240/260 μm double truncated circular inner cladding, with a low index polymer jacket as the outer cladding. The NA is 0.12 and 0.46 for the core and inner cladding, respectively. This step index fiber supports single transverse mode operation for wavelengths longer than 2.5 μm. The fiber output tip was cleaved at a 4 degrees angle and the light emerging from the fiber tip was collimated using an anti-reflective coated ZnSe asphere (BAE Australia). The cavity was completed by a diffraction grating blazed for 3.5 μm (Thorlabs GR2550-30035). The zero order of the grating was used as the output coupler of the fiber laser. A silver mirror was mounted at 90 degrees to the grating on the same mirror mount which prevented the output beam from changing its pointing angle during tuning at the cost of a small transverse movement. This effectively acted as a corner reflector which considerably simplified the monitoring of the laser output.

The laser tuning range for three different incident power levels of $P_2$ is presented in Figure 4. In these experiments the incident power of $P_1$ was maintained at 5 W (Higher $P_1$ power did not increase the laser output power indicating that $P_2$ absorption was saturated). Increasing $P_2$ to 6 W increased the output power to 100 mW at the center of the tuning curve at 3550 nm. However, the laser could not sustain this power level and would cease operation after few minutes due to damage to the fiber core. The use of fiber end-capping [10] would alleviate this issue. During operation with $P_2$ at lower power levels of 2 W and 3 W, some residual 2.8 µm lasing (5 mW and 2 mW, respectively) was observed. The 2.8 µm power was constant regardless of the output power or wavelength of operation at the 3.5 µm band. Tunable laser power was lower than the power obtained with a simple open cavity configuration when the grating is replaced with an 80% reflectivity output coupler. This suggests that the current configuration could be optimized with use of a higher efficiency grating and reducing the losses in the collimating optics.

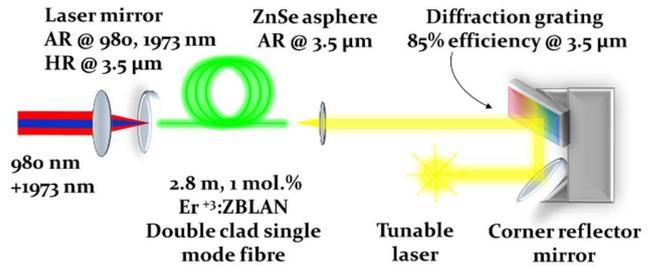

**Figure 3** – Schematic diagram of the tunable fiber laser

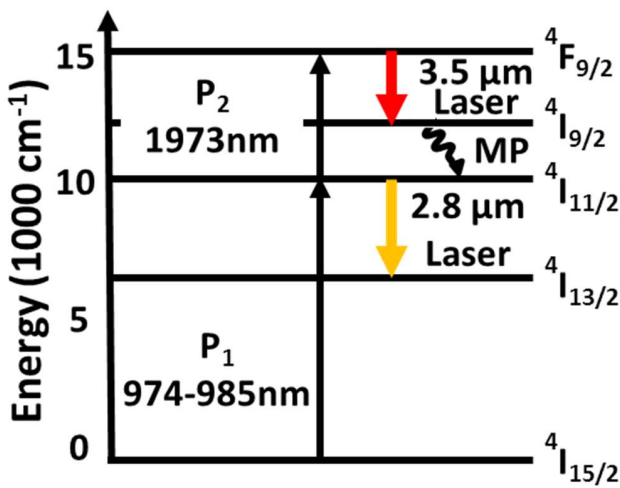

**Figure 2** – Simplified energy level diagram showing the laser transitions and the dual-wavelength pumping process.

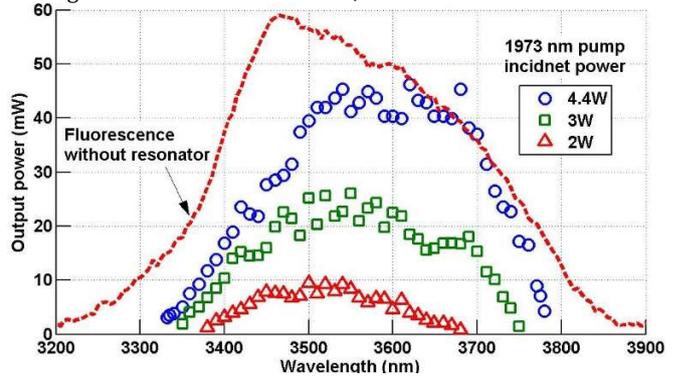

**Figure 4** – Measured tuning characteristics of the fiber laser using a pump power of 5 W at 977 nm and different levels of 1973 nm pump power. The red dashed line shows the fluorescence spectrum obtained from the $^4F_{9/2} \rightarrow {}^4I_{9/2}$ transition without the resonator in place.

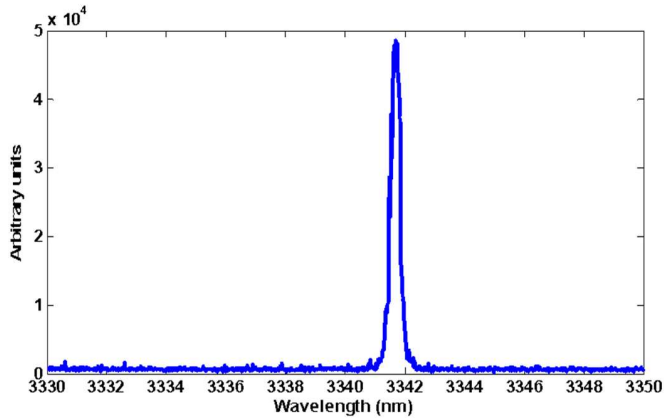

Figure 5 - Typical output laser linewidth of the tunable laser.

The laser linewidth observed throughout the 3.5 µm tuning band was 0.3 nm (Figure 5), a little over twice the monochromator minimum resolution of 0.13 nm indicating that the laser was not running on a single longitudinal mode. At a maximum tuned power of 47 mW, this represents a spectral brightness of 160mW/nm which compares favorably with recent reports of supercontinuum generation in this region of the mid-infrared [11]. The free-spectral range of this resonator is 32 MHz and mode beating was observed at this wavelength using an RF spectrum analyzer. Single longitudinal mode operation might be possible in the future using additional dispersive elements in the resonator or a DBR approach with a fiber Bragg gratings [12] or by adding additional frequency discrimination using intra-cavity etalons.

For a fixed amount of Stark splitting of the lower laser level, the maximum tuning range of a tunable laser system should scale with the centre wavelength of the emission. Thus the ratio $\Delta E_L/E_{ZL}$, where $\Delta E_L$ is the total Stark splitting of the lower laser level and $E_{ZL}$ is the zero line energy for the transition, would increase as the center wavelength increases (i.e., $E_{ZL}$ decreases). This trend is somewhat borne out from demonstrations of widely tunable rare earth doped fiber lasers that have involved the ground-state terminated transitions of $Yb^{3+}$ [13], $Er^{3+}$ [14] and $Tm^{3+}$ [15] in silicate glasses, that have displayed tuning ranges of 144 nm, 100 nm and 255 nm respectively. (These tuning ranges provide $\Delta\lambda_r/\lambda_c$ ratios, where $\Delta\lambda_r$ is the tuning range and $\lambda_c$ is the centre wavelength of the tuning of 0.14, 0.063 and 0.13, respectively.) For our system, calculations using the Stark level assignments for $Er^{3+}$ in a crystalline host [16] give $\Delta E_L/E_{ZL}$ = 0.15, which equates to a potential tuning range of 0.53 µm. Our measured tuning range of 450 nm is 83% of this value and shows that the tuning could be extended with more pumping at 1976 nm.

Laser operation at 2.8 µm could be achieved by switching off the $P_2$ pump and tuning the grating such that the wavelength of the $^4I_{11/2} \rightarrow {}^4I_{13/2}$ transition was selected. These results are illustrated in Figure 6. This extends the coverage of this laser such that it overlaps absorption lines of carbon dioxide and ammonia for example. The performance of this laser transition is currently limited due to the lack of optimal cavity mirrors and collimating aspheres. In addition, the gain medium is lightly doped with erbium and hence bottlenecking in the lower lasing state is likely at these power levels limiting the performance and tunability characteristics of this transition. (Co-doping with $Pr^{3+}$ ions should also widen the tuning without significantly interfering with the performance of the mid-infrared laser.)

We investigated the power scaling potential using dielectric mirrors only which employed the same experimental layout as shown in Figure 3, except the output facet was re-cleaved perpendicular and the intra-cavity aspheric and diffraction grating were replaced with an output-coupler (OC) with 80% reflectivity between 3.4 µm and 3.9 µm. The beam emerging from the laser resonator was collimated using an anti-reflective coated ZnSe asphere (BAE Australia). The outgoing laser and the residual pumps were separated to monitor the residual pump power while the laser output was split to allow simultaneous measurement of the power (with a thermal power meter), wavelength (using a grating monochromator) and beam profile (using a thermal camera). Such a configuration allowed us to observe possible 2.8 µm lasing and the change in the wavelength of operation of the 3.5 µm laser.

Using 2 W of $P_1$ and a little over 4 W of $P_2$ produced 0.9 W at 3.47 µm with a slope efficiency of 27% relative to incident $P_2$ (figure 7). A maximum power of 1.45 W was achieved with $P_1$ operating at 4 W and $P_2$ at a maximum of 5.5 W. This corresponds to an overall optical efficiency of 15%. At this power level the laser output suffered from strong instabilities, possibly due to thermal and mechanical stress at the fiber pump input facet, which prevented additional pump power from increasing the laser power further. It is clear that the recent demonstration [10] that employs fiber post-processing to avoid large intensities at the glass-air boundary is necessary at elevated power levels. The laser operated on a number of lines in the 3.45 μm to 3.47 μm region with the distribution of power shifting to the longer lines with increasing pump power (Figure 8). Wavelength stability was reduced at the higher power levels with multiple lines competing. The laser output beam was observed to be near diffraction limited with an $M^2$ of 1.08 ±0.3 at an output power of 1 W (see Figure 9).

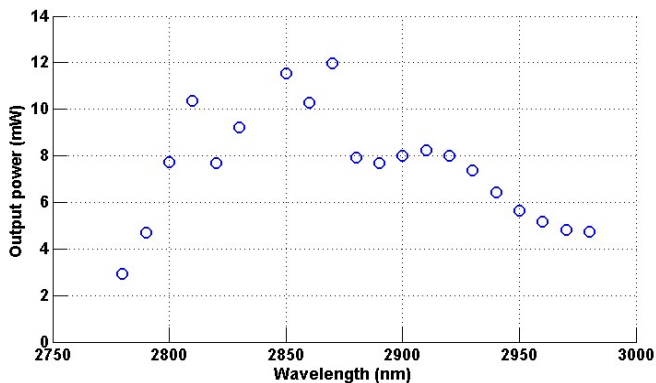

Figure 6 – Measured tuning characteristic of the 2.8 m long fiber laser when operated on the $^4I_{11/2} \rightarrow {}^4I_{13/2}$ transition at 2.8 µm. 6 W of $P_1$ only is incident on the fiber.

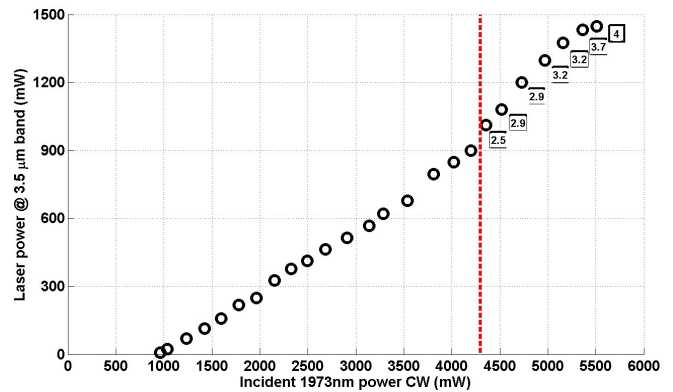

Figure 7 – Measured output power as a function of pumping with $P_2$ for the dielectric mirror only resonator. $P_1$ was maintained at 2 W up to the red line, beyond which it was increased incrementally (see numbers to the right of data) to a maximum of 4 W. Instabilities in the laser power prevented obtaining higher output power.

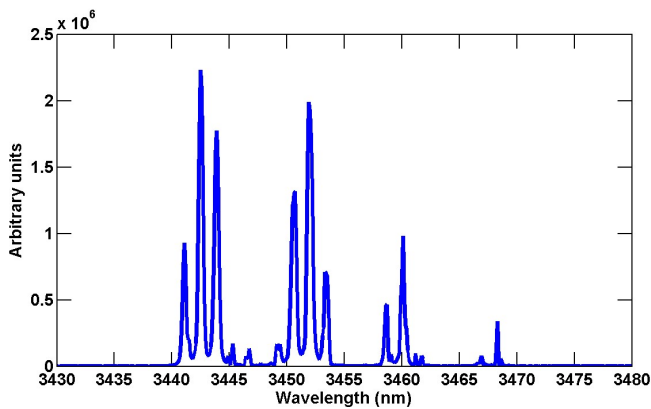

**Figure 8 – Close up of typical laser lines running at higher power levels. With increasing output power the lines would redistribute the power to operate at longer wavelengths.**

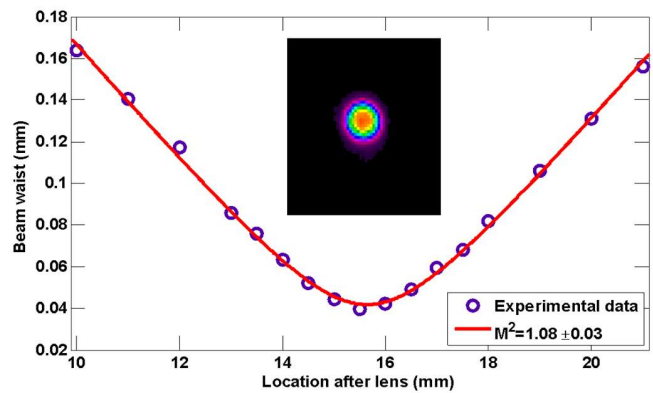

**Figure 9 - Typical output laser beam quality and beam shape (inset) at 1 W of output power.**

In this work we have shown the extension of the dual-wavelength pumping concept to the development of a broadly tunable mid-infrared fiber laser. The maximum tuning range was 450 nm, the widest from any fiber laser and compliments the emission characteristics from quantum cascade lasers, albeit with better beam quality and better power scalability potential. In a simple dielectric mirror-defined resonator, the system produced 1.45 W of output power at a wavelength of 3.47 um and slope efficiency of 27%, making this setup one of the most efficient in this region of the mid-infrared.

Dr. Henderson-Sapir and Dr. Ottaway gratefully acknowledge the financial support of the University of Adelaide CAS grant and The South Australian Government Premiers Research International Fund (PRIF). Dr. Jackson acknowledges funding through the Australian Research Council (ARC) Discovery Projects scheme.

We sincerely thank Prof. Jesper Munch for his useful suggestions, Mr. Andrew Malouf for his help in the lab and Mr. Robert Chivell for his assistance with the diffraction grating mount.